\documentclass[aps,nofootinbib,preprint]{revtex4}
\usepackage{amsfonts,amssymb,amsmath}
\usepackage{graphicx}
\usepackage{subfigure}
\usepackage{wrapft}
\def\d{\delta}

\def\nt{\notag}

\newcommand{\f}[2]{\frac{#1}{#2}}
\newcommand{\mk}[1]{\left( #1 \right)}
\newcommand{\kk}[1]{\left[ #1 \right]}

\newcommand{\be}{\begin{equation}}
\newcommand{\ee}{\end{equation}}
\newcommand{\beq}{\begin{equation}}
\newcommand{\eeq}{\end{equation}}
\newcommand{\F}{{\cal{F}}}
\newcommand{\lmk}{\left(}  \newcommand{\rmk}{\right)}

\newcommand{\vecs}[1]{\mbox{\boldmath\tiny ${#1}$}}
\newcommand{\veck}{\vecs k}
\newcommand{\keff}{k_{\rm eff}}
\newcommand{\lcdm}{{\rm \Lambda CDM}}
\newcommand{\frg}{{\rm fRG}}
\newcommand{\eff}{\text{eff}}
\begin{document}

\title{
Phantom boundary crossing and anomalous growth index of
fluctuations in viable $f(R)$ models of cosmic acceleration
}

\author{
Hayato Motohashi$^{~1,2}$, Alexei A. Starobinsky$^{~2,3}$,
and Jun'ichi Yokoyama$^{~2,4}$%
}

\address{
$^{1}$ Department of Physics, Graduate School of Science,
The University of Tokyo, Tokyo 113-0033, Japan \\
$^{2}$ Research Center for the Early Universe (RESCEU), 
Graduate School of Science, The University of Tokyo, Tokyo 113-0033, Japan \\
$^{3}$ L. D. Landau Institute for Theoretical Physics, 
Moscow 119334, Russia \\
$^{4}$ Institute for the Physics and Mathematics of the Universe(IPMU), 
The University of Tokyo, Kashiwa, Chiba, 277-8568, Japan
}

\begin{abstract}
Evolution of a background space-time metric and sub-horizon matter
density perturbations in the Universe is numerically analyzed in
viable $f(R)$ models of present dark energy and cosmic
acceleration. It is found that viable models generically exhibit
recent crossing of the phantom boundary $w_{\rm DE}=-1$. Furthermore, it is
shown that, as a consequence of the anomalous growth of density
perturbations during the end of the matter-dominated stage, their
growth index evolves non-monotonically with time and may even become negative
temporarily.
\end{abstract}

\begin{flushright}
RESCEU-5/10
\end{flushright}

\maketitle

\section{Introduction}

The physical origin of the dark energy (DE) which is responsible
for an accelerated expansion of the current Universe is one of the
largest mysteries not only in cosmology but also in fundamental
physics \cite{review}. Although the standard spatially flat
${\rm \Lambda}$-Cold-Dark-Matter ($\lcdm$) model is consistent with
all kinds of current observational data \cite{WMAP7},
some tentative deviations from it have been reported recently
\cite{Shafieloo:2009ti,Bean:2009wj} which, if proven to be not due
to systematic and other errors, may eventually rule out an exact
cosmological constant. Furthermore, in the $\lcdm$ model, the
cosmological term is regarded as a new fundamental constant whose
observed value is much smaller than any other energy scale known
in physics. So, its understanding in fundamental physics  is
lacking today, although some non-perturbative effects may generate
such a small quantity \cite{Yokoyama:2001ez}. On the other hand,
we know that ``primordial DE,'' which is responsible for inflation
in the early universe \cite{S80,sato,guth}, is not identical to
the cosmological constant, in particular, it is not stable and
eternal. Hence it is natural to seek for non-stationary models of
the current DE, too.

Among them, $f(R)$ gravity which modifies and generalizes the
Einstein gravity by incorporating a new phenomenological function
of the Ricci scalar $R$, $f(R)$, provides a self-consistent and
non-trivial alternative to $\lcdm$ model, see {\it e.g.} Ref.\
\cite{SF08} for a recent review. This theory is a special class
of the scalar-tensor theory of gravity with the vanishing
Brans-Dicke parameter $\omega_{BD}$
\cite{Chiba:2003ir,Tsujikawa:2008uc}. It contains a new scalar
degree of freedom dubbed "scalaron" in Ref.~\cite{S80}, thus, it is a
{\em non-perturbative} generalization of the Einstein gravity.

This additional degree of freedom imposes  a number of conditions
on viable functional forms of $f(R)$. In particular, in order to
have the correct Newtonian limit for $R\gg R_0\equiv R(t_0)\sim
H_0^2$ where $t_0$ is the present moment and $H_0$ is the Hubble
constant, as well as the standard matter-dominated stage with the
scale factor behaviour $a(t)\propto t^{2/3}$ driven by cold dark
matter and baryons, the following conditions should be fulfilled:
\beq |f(R)-R|\ll R,~~|f'(R)-1|\ll 1,~~Rf''(R)\ll 1, ~~R\gg R_0~,
\label{ineq} \eeq 
where the prime denotes the derivative with
respect to the argument $R$. In addition, the stability condition
$f''(R)>0$ has to be satisfied that guarantees that the standard
matter-dominated Friedmann stage remains an attractor with respect
to an open set of neighboring isotropic cosmological solutions in
$f(R)$ gravity. In quantum language, this condition means that
scalaron is not a tachyon.  Note that the other stability
condition, $f'(R)>0$, which means that gravity is attractive and
graviton is not a ghost, is automatically fulfilled in this
regime. Specific functional forms that satisfy all these
conditions have been proposed in 
Refs.~\cite{Hu:2007nk,AB07,Starobinsky:2007hu} etc., 
and much work has been done on their cosmological consequences.

In the previous paper \cite{Motohashi:2009qn} we calculated
evolution of matter density fluctuations in viable $f(R)$ models
\cite{Hu:2007nk,Starobinsky:2007hu} in the limiting case $R\gg
R_0$ during the matter-dominated stage and found an analytic
expression for them. In this paper we extend the previous analysis
and perform numerical calculations of the evolution of both
background space-time and density fluctuations for the particular
$f(R)$ model of Ref.~\cite{Starobinsky:2007hu} without such
restriction on $R$. As a result, we have found the phantom
boundary crossing at an intermediate redshift $z\lesssim 1$ for the 
background space-time metric and an anomalous behaviour of the growth 
index of fluctuations.

The rest of the paper is organized as follows.  In \S 2 we
introduce evolution equations for the homogeneous and isotropic
background and present results of numerical integration. In \S 3
we report numerical solutions for the evolution of density
fluctuations and other observables. Section 4 is devoted to
conclusions and discussion.

\section{Evolution of the background Universe}

We adopt the following action with a four-parameter family of
$f(R)$ models:
\begin{align}
S&= \frac{1}{16\pi G} \int d^4x \sqrt{-g} f(R) + S_m, \label{S}\\
f(R)&=R + \lambda R_s \left[ \left( 1 + \frac{R^2}{R_s^2}\right)^{-n}
-1
\right] +\frac{R^2}{6M^2}, \label{fR}
\end{align}
where $n,~\lambda,~R_s$, and $M$ are model parameters and $S_m$ is
the action of the matter content which is assumed to be minimally
coupled to gravity (thus, the action \eqref{S} is written in the
Jordan frame). This is the model of Ref.\
\cite{Starobinsky:2007hu} modified by the last term in \eqref{fR}
borrowed from the inflationary model of Ref.~\cite{S80}. 
This term is introduced for several purposes associated with
high-curvature behaviour of the theory. One of them, as explained
in Ref.~\cite{Starobinsky:2007hu} , is to avoid excessive growth of the
scalaron mass, $m_s^2=1/3f''(R)$ in the regime \eqref{ineq},
towards the early Universe, $t\to 0$. The other one is to remove
the additional and undesirable ``Big Boost'' singularity which can
arise in the original models
\cite{Hu:2007nk,AB07,Starobinsky:2007hu} as was shown in Ref.\
\cite{F08} (see Refs.~\cite{Motohashi:2009qn,TSC09,ABS09} for more
discussion on this point). The value of $M$ should be sufficiently
large in order not to destroy the standard cosmology of the
present and early Universe. In particular, the values of $M$
considered in Refs.~\cite{D08,KM09} are not high enough for this
purpose, because $M$ should not be smaller than the Hubble
parameter $H(t)$ during the $N\sim 60$ last e-folds of inflation
in the early Universe in order to avoid overproduction of relic
scalarons, as well as to solve other cosmological problems. In
fact, if we take $M\approx 3\times 10^{13}$ GeV, the scalaron
itself can act as an inflaton \cite{S80} and generate primordial
scalar (adiabatic) and tensor perturbations \cite{MC81,S83} with
the amplitudes and slopes of their power spectra in agreement with
all observational data available today. Note, however, that as
shown in Ref.~\cite{ABS09}, such a "unified" model describing both
primordial DE driving inflation in the early Universe and present
DE driving recent acceleration of the Universe in the scope of
$f(R)$ gravity leads to slightly different predictions for
parameters of the primordial perturbation spectra, as compared to
the purely inflationary model with $\lambda R_s=0$, due to a
change in the number of observable e-folds of inflation $N$ caused
by different evolution of the Universe during generation and
heating of usual matter after inflation. Furthermore, in this
unified model the term in the square brackets in \eqref{fR} should
be modified for $|R|<R_0$ in such a way as to ensure the
fulfillment of the stability condition $f''(R)>0$ in this region,
too.

So, we take this value of $M$ and assume that the evolution of the
Universe is identical to that in the standard $\lcdm$ model
at high redshifts without any relic scalaron oscillations. Then
the $R^2/6M^2$ term is totally negligible in the epoch we are
concerned here. Therefore, we do not include its contribution
below.

We can express field equations derived from the action in the
following Einsteinian form. 
\beq
 R^{\mu}_{\nu}-\frac{1}{2}\delta^{\mu}_{\nu}R=
-8\pi G\lmk T^{\mu}_{\nu (m)}+T^\mu_{\nu ({\rm DE})}\rmk, \eeq 
where
\beq 8\pi G T^\mu_{\nu ({\rm DE})}\equiv
\F'(R)R^\mu_\nu-\frac{1}{2}\F(R)\delta^{\mu}_{\nu}
+(\nabla^\mu\nabla_\nu-\delta^{\mu}_{\nu}\square)\F'(R),~~~
\F(R)\equiv f(R)-R  \label{EMtensor} \eeq 
(the sign conventions
here are the same as in Ref.~\cite{Starobinsky:2007hu}). Working in the
spatially flat Friedmann-Robertson-Walker (FRW) space-time with
the scale factor $a(t)$, we find
\begin{align}
3H^2&=8\pi G\rho-3\F' H^2+\frac{1}{2}(\F' R-\F)-3H\dot{\F}',\label{hubble}\\
2\dot{H}&=-8\pi G\rho -2\F'\dot{H}-\ddot{\F}'+H\dot{\F}',\label{hdot}
\end{align}
where $H$ is the Hubble parameter and $\rho$ is the energy density
of the material content which we assume to consist of
non-relativistic matter.

From \eqref{EMtensor} the
effective energy density and pressure of dark energy can be expressed as
\begin{align}
&8\pi G\rho_{\rm DE}=\frac{1}{2}(\F'R-\F)-3H^2\F'-3H\dot{\F}'
=-3H\dot{R}\F'' +3(H^2+\dot{H})\F'-\frac{1}{2}\F,\label{rhoDE}\\
&8\pi G(\rho_{\rm DE}+P_{\rm DE})=2\dot{H}\F'-H\dot{\F}'+\ddot{\F}', \label{PDE}
\end{align}
respectively, where $R=12H^2+6\dot{H}$. We define the DE equation
of state parameter $w_{\rm DE}$ by the ratio $w_{\rm DE}\equiv
P_{\rm DE}/\rho_{\rm DE}$.

With the appropriate initial condition after cosmic inflation
mentioned above, $\F$ takes an asymptotically constant value
$\F=-\lambda R_s$ at high redshift (apart from the $R^2/6M^2$ term
which we neglect here).  In this regime, evolution of the Universe
is the same as that obtained from the Einstein action with a
cosmological constant $\Lambda(\infty)=\lambda R_s/2$. The scale
factor therefore evolves as 
\beq a=a_i\lmk\frac{16\pi G\rho_i}{\lambda R_s}\rmk^{\frac{1}{3}}
\sinh^{\frac{2}{3}}\lmk\sqrt{\frac{3\lambda R_s}{8}}t\rmk \cong
a_i\lmk\frac{t}{t_i}\rmk^{\frac{2}{3}}, \eeq 
where the suffix $i$ denotes quantities at an initial time $t=t_i$.

The time dependence of $\rho_{\rm DE}$ is mainly governed by the first
term in the right-most expression of \eqref{rhoDE} initially.
Since $\dot{R}< 0$ and $\F''>0$ for stability, this means that the
effective energy density of dark energy {\it increases} with time
in this regime. Therefore, DE exhibits the phantom behaviour,
$w_{\rm DE}<-1$, during the matter-dominated stage with $z>1$, which
lasts only temporarily because the late-time asymptotic de Sitter
stage has an effective cosmological constant smaller than
$\Lambda(\infty)$. So, $\rho_{\rm DE}$ stops growing after the end of
the matter-dominated stage and begins to decrease.

Indeed, as shown in Ref.\ \cite{Starobinsky:2007hu}, the late-time 
asymptotic de Sitter
solution has a curvature $R\equiv R_1\equiv x_1R_s$ where $x_1$
is the maximal solution of the equation,
\beq   \lambda = \frac{x(1+x^2)^{n+1}}{2\kk{(1+x^2)^{n+1}-1-(n+1)x^2}}. \eeq 
It satisfies the inequality $x_1<2\lambda$, so that
$\Lambda(R_1)=R_1/4<\Lambda(\infty)$. These inequalities are
saturated in the limit $n \gg 1$ for fixed $x_1$, or $x_1 \gg 1$
for fixed $n$.  In these cases cosmic evolution is
indistinguishable from the standard $\lcdm$ model.

Thus, this model naturally realizes crossing of the phantom
boundary $w_{\rm DE}=-1$ in a recent epoch. Note that phantom
behaviour of DE is generic in its models based on the
scalar-tensor gravity \cite{BEPS00} which includes the $f(R)$
theory. Here we see that it is realized in all simplest stable
$f(R)$ models of present DE.

The stability condition of this future de Sitter solution\cite{MSS88} ,
$f'(R_1)>R_1f''(R_1)$, imposes the following constraint on $x_1$.
\beq  (1+x_1^2)^{n+2} > 1 + (n+2)x_1^2 + (n+1)(2n+1)x_1^4,
\label{x1constraint} \eeq 
which is stronger than any other
constraint discussed above. For each $n$ we can find $x_1$ which
marginally satisfies \eqref{x1constraint} and gives the minimal
allowed value of $\lambda$.  Numerically we find
$(n,x_{1\min},\lambda_{\min})=(2,1.267,0.9440),(3,1.041,0.7259)$,
and $(4,0.9032,0.6081)$ for each $n$, respectively (if $n=2$, the 
analytic expression for $x_{1\min}$ is $x_{1\min}^2=\sqrt{13} - 2$). 
For comparison, the analytic results for $n=1$ are $x_{1\min}=\sqrt
3\approx 1.732,~\lambda_{\min}=8/(3\sqrt3)\approx 1.540$.

We numerically solve evolution equation \eqref{hdot} using
\eqref{hubble} to check numerical accuracy, taking $t_i$
at the epoch when matter density parameter took
$\Omega_i=16\pi G\rho_i/(16\pi G\rho_i+\lambda R_s)=0.998$.
We determine the current epoch  by the requirement that
the value of $\Omega$ takes the observed central value
$\Omega_0=0.27$ and  $R_s$ is fixed so that
the current Hubble parameter
$H_0=72$km/s/Mpc is reproduced.
We find  the ratio $R_s/H_0^2$ is well fit by a simple
power-law $R_s/H_0^2=c_n\lambda^{-p_n}$ with
$(n,c_n,p_n)=(2,4.16,0.953),~(3,4.12,0.837),$ and $(4,4.74,0.702)$,
respectively, whereas in the $\lcdm$ limit it would behave as
$R_s/H_0^2=6(1-\Omega_0)/\lambda\simeq 4.38\lambda^{-1}$.

\begin{figure}[t]
\centering
\subfigure[Evolution of $w_{\rm DE}$ for different values of $\lambda$
 with $n=2$.]{
\includegraphics[width=77mm]{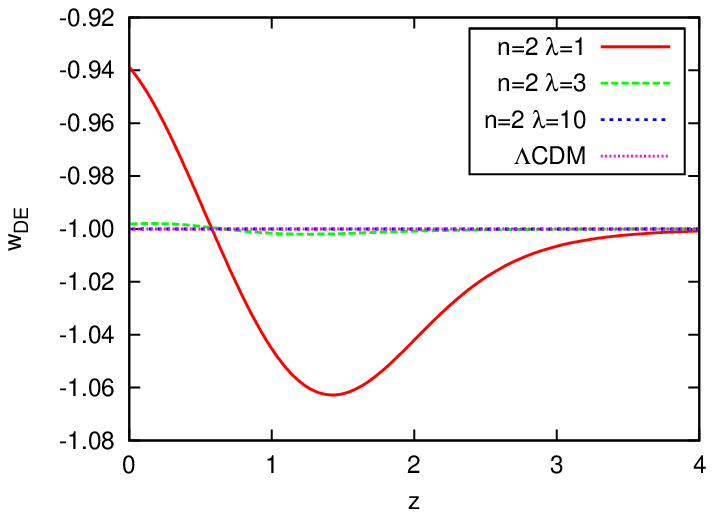}
\label{fig:1a}}
\subfigure[Evolution of $w_{\rm DE}$ for $\lambda_{\min}$
for $n=2,3,$ and 4.]{
\includegraphics[width=77mm]{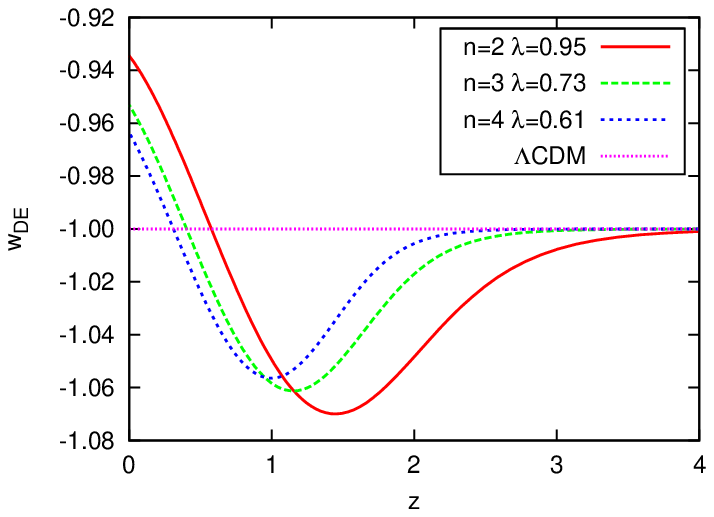}
\label{fig:1b}}
\caption{Evolution of the equation-of-state parameter of
effective dark energy.}
\label{fig:outbg_w}
\end{figure}

Figures \ref{fig:outbg_w} depict evolution of $w_{\rm DE}$ as a function of redshift
$z$ where phantom crossing is manifest. As expected, it approaches
$w_{\rm DE}=-1=\text{constant}$ as we increase $\lambda$ for fixed
$n$. For minimal allowed values of $\lambda$, deviations from
$w_{\rm DE}=-1$ are observed at $\sim  5\%$ level in both directions for $z
\lesssim 2$ independently of $n$. 
Such behaviour of $w_{\rm DE}$ is well admitted by all most recent
observational data, see e.g. Ref.~\cite{WMAP7}. The average value
of $w_{\rm DE}$ over the interval $0 \le z \le 1$ to which all BAO and most
of SN data refer is very close to $-1$. Moreover, in this range (but not
for larger values of $z$), the behaviour of $w_{\rm DE}$ for minimal
allowed values of $\lambda$ (i.e. for largest possible deviations
from the $\lcdm$ background model) is well fitted by the CPL fit\cite{CPL}
$w_{\rm DE}(z)=w_0+w_az/(1+z)$ with 
$(n,\lambda_{\rm min},w_0,w_a)=$
$(2,0.95,-0.92,-0.23)$, $(3,0.73,-0.94,-0.22)$ 
and $(4,0.61,-0.96,-0.21)$, respectively. 
$|1+w_0|$ and $|w_a|$ decrease slowly
for larger values of $n$. These values of $w_0$ and $w_a$
lie very close to the center of the $68\%$ and $95\%$ CL ellipses
for all combined data in Fig. 13 of Ref.~\cite{WMAP7}.

As explained above, this phantom
crossing behaviour is not peculiar to the specific choice of the
function \eqref{fR} but a generic one in models which satisfy the
stability condition $\F''>0$. Indeed, a similar behaviour has been
observed in other $f(R)$ DE models, too
\cite{Martinelli:2009ek,ABS09}. We also note that different
definitions of $\rho_{\rm DE}$, $P_{\rm DE}$, and $w_{\rm DE}$ have been used
in literature \cite{Amendola:2007nt} which lead to different
behaviour of $w_{\rm DE}$.

Although the behaviour of dark energy is quite different depending
on model parameters, the total expansion factor $a_0/a_i$
from the epoch $\Omega_i=0.998$ to the present varies only
between $a_0/a_i=10.8$ and 11, the latter corresponding to the
value in the $\lcdm$ model.

We have also calculated the quantity
$B(z) = (f''/f')(dR/d\ln H)$ introduced in Ref.~\cite{Song:2006ej} at present time.
We have found $B(0)=0.21$, $6.1\times 10^{-5}$, and $0.17$, 
for $(n,\lambda)=(2,0.95)$, $(2,8)$, and $(4,0.61)$, respectively. 

\section{Density fluctuations}

We now turn to evolution of density fluctuations.
In $f(R)$ gravity, the evolution equation of density fluctuations,
$\delta$, deeply in the sub-horizon regime is given by
\cite{Zhang:2005vt,Tsujikawa:2007gd}
\beq \label{de1} \ddot \d + 2H\dot \d - 4\pi G_\eff \rho \d = 0, \eeq
where
\be \label{tsuji} G_{\text{eff}}=\frac{G}{F}
\frac{1+4\frac{k^2}{a^2}\frac{F'}{F}}
{1+3\frac{k^2}{a^2}\frac{F'}{F}},~~~F(R)\equiv f'(R). \ee
This equation reduces to the correct evolution equation
for all wavenumbers for the CDM model in the Einstein gravity where $F=1$.

In the previous paper\cite{Motohashi:2009qn} we obtained
an analytic solution in the high-curvature regime
when the scale factor evolves as $a(t)\propto t^{2/3}$
and $F$ takes the asymptotic form
\beq
  F\simeq 1-2n\lambda \mk{\f{R}{R_s}}^{-2n-1}\equiv 
  1-\mk{\f{R}{R_c}}^{-N-1}, \label{kinjiF}
\eeq
with the following correspondence:
\beq
N=2n ~~~{\rm and}~~~
R_c=R_s(2n\lambda)^{1/(2n+1)}.
\eeq
The two independent solutions of \eqref{de1} in this regime read
\begin{align}
&\delta_{\veck}(t)=\delta_{i\veck}\left(\frac{t}{t_i}\right)^{\frac{-1\pm
5}{6}} \nonumber \\
&\times \,_2F_1\left(\frac{\pm 5-\sqrt{33}}{4(3N+4)}, \frac{\pm
5+\sqrt{33}}{4(3N+4)};1\pm\frac{5}{2(3N+4)};
-3\frac{(N+1)k^2}{a_i^2R_c^2}\left(\frac{t}{t_i}\right)^{2N+8/3}
\right)  \label{as}
\end{align}
in terms of the hypergeometric function\cite{Motohashi:2009qn}.
In the following discussion, we consider the upper
sign solution only, because the other solution corresponds to the
decaying mode  and is singular at $t\to 0$.  Then the solution behaves
as
\beq
  \delta_{\veck}(t)\xrightarrow[]{t\to 0} \delta_{i\veck}
\left(\frac{t}{t_i}\right)^{\frac{2}{3}}
~~{\rm and}~~
\delta_{\veck}(t)\xrightarrow[]{t\to \infty}
\delta_{i\veck}C(k)\left(\frac{t}{t_i}\right)^{\frac{-1+\sqrt{33}}{6}},
 \label{limiting}
\eeq
respectively.  The transfer function, $C(k)$, is given by
\begin{align}
C(k)&=\f{\Gamma\left(1+\frac{5}{2(3N+4)}\right)
\Gamma\left(\frac{\sqrt{33}}{2(3N+4)}\right)}
{\Gamma\left(1+\frac{5+\sqrt{33}}{4(3N+4)}\right)
\Gamma\left(\frac{5+\sqrt{33}}{4(3N+4)}\right)}
\left[\frac{3(N+1)k^2}{a_i^2R_c}
\left(\frac{3R_ct_i^2}{4}\right)^{N+2}\right]
^{\frac{-5+\sqrt{33}}{4(3N+4)}} \nt \\
&= \f{ \Gamma\mk{1+\frac{5}{4(3n+2)}}\Gamma
\mk{\frac{\sqrt{33}}{4(3n+2)}} }
{ \Gamma\mk{1+\frac{5+\sqrt{33}}{8(3n+2)}}
\Gamma\mk{\frac{5+\sqrt{33}}{8(3n+2)}} }
\kk{\f{6n\lambda (2n+1)k^2}{a_i^2 R_s}
\mk{\frac{3R_st_i^2}{4}}^{2(n+1)}}^{\f{-5+\sqrt{33}}{8(3n+2)}},
\label{transfer}
\end{align}
where
\beq
  t_i=\f{2}{3}\sqrt{\f{6}{\lambda R_s}}\sinh^{-1}
\sqrt{\f{1-\Omega_i}{\Omega_i}}.
\eeq

Note that the effective gravitational constant \eqref{tsuji} reads
\beq
G_{\rm eff}=G\mk{1+\f{1}{3}\f{k^2/a^2m_s^2}{1+k^2/a^2m_s^2}},
\eeq
in the high-curvature regime when $F\cong 1$.  In the position space,
such a theory has the potential
\beq
  V(r)=-\f{G}{r}\mk{1+\f{1}{3}e^{-m_sr}},
\eeq
per unit mass \cite{Gannouji:2008wt} for such sufficiently small $r$ for which 
time dependence of $m_s(t)$ may be neglected.
Thus, each Fourier mode feels $4/3$ times the conventional
gravitational force if and only if
 $k/a(t)\gtrsim m_s(t)=\mk{3F'}^{-1/2}$.

The transition from former temporal behaviour to the latter one in
\eqref{limiting} occurs at the epoch $t_k$ determined by
\beq
 k=a(t_k)m_s(t_k)=a(t_k)\mk{\frac{R_s}{6n(2n+1)\lambda}}^{\f{1}{2}}
\mk{\f{R(t_k)}{R_s}}^{n+1}.  \label{kprop}
\eeq
The above expression is proportional to $ t_k^{-2n-4/3}$ for
those modes which physical wavenumber (momentum) $k/a(t)$ crosses the scalaron 
mass $m_s(t)$ in the high-curvature regime. This explains
$k$-dependence of the transfer
function \eqref{transfer}\cite{Starobinsky:2007hu}.
If we adopt an expression of $R(t)$ in $\lcdm$,
\beq
  R(t)=3H_0^2\kk{\Omega_{m0}\mk{\f{a_0}{a(t)}}^3+4(1-\Omega_{m0})},
\label{rt}
\eeq
we can further approximately obtain the crossing time,
$t_{\ast}(k)$,
 for a smaller wavenumber, $k_{\ast}$, as well:
\beq
  \f{k_{\ast}}{a(t_{\ast})}=
\f{\lambda^{\mk{n+\f{1}{2}}p_n-\f{1}{2}}}{\sqrt{6n(2n+1)}c_n^{n+\f{1}{2}}}
\kk{3\Omega_{m0}\mk{\f{a_0}{a(t_{\ast})}}^3+12(1-\Omega_{m0})}^{n+1}H_0.
\label{kscalaron}
\eeq
From \eqref{kscalaron} we find that the physical wavenumber crossing the 
scalaron mass today is given by
\beq
\label{pwcsmt} \f{k_0}{a_0}=\frac{9.57^{n+1}\lambda^{\mk{n+\f{1}{2}}p_n-\f{1}{2}}}
{\sqrt{6n(2n+1)}c_n^{n+\f{1}{2}}}H_0=
\begin{cases}
3.2\lambda^{1.88}H_0 & (n=2)\\
5.3\lambda^{2.43}H_0 & (n=3)~.\\
5.0\lambda^{2.66}H_0 & (n=4)
\end{cases}
\eeq
Thus, except for cases with large  $\lambda$, all observable scale feels
the scalaron force today.

\begin{figure}[t]
\centering
\includegraphics[width=140mm]{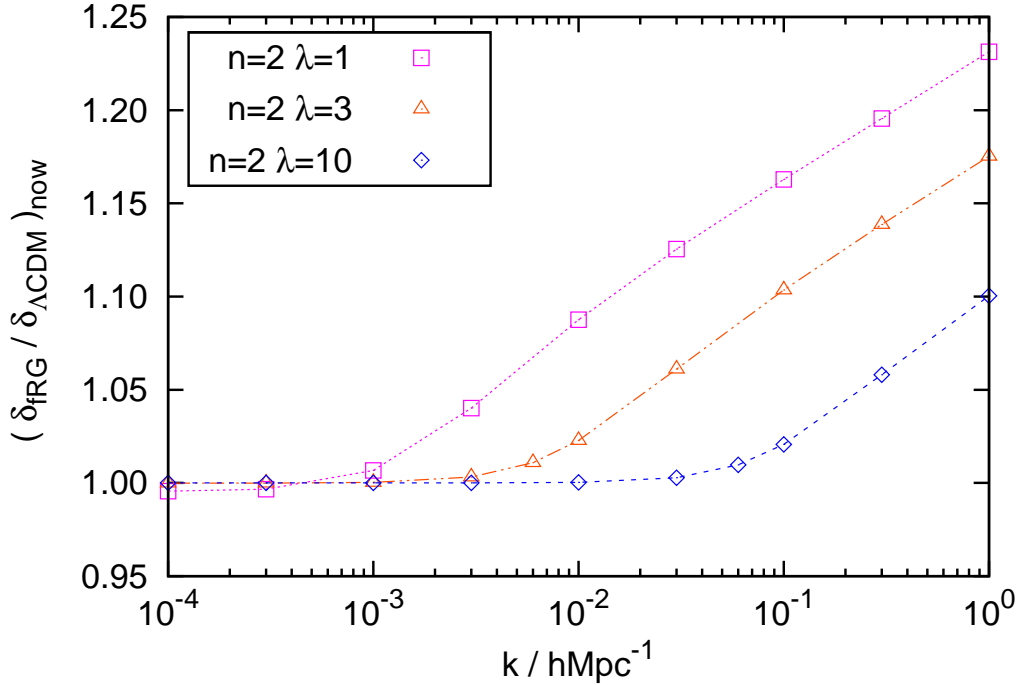}
\caption{ The ratio of linear density perturbations
$\delta_{\frg}/\delta_{\lcdm}$ at present as a
 function of $k$ for three different values of $\lambda$ with $n=2$.}
\label{fig:dk_n=2_l}
\end{figure}

Since the analytic solution \eqref{as} is valid in the high-curvature era
only, we must solve \eqref{de1} numerically to obtain a
full solution using the analytic solution as an initial condition.
Figure \ref{fig:dk_n=2_l} depicts the ratio of linear density fluctuation in $f(R)$
model, $\delta_{\frg}$, to that in the $\lcdm$ model,
$\delta_{\lcdm}$, with the same initial condition. 
Fluctuations with small wavenumbers have practically the same value as those in
the $\lcdm$ model, while those on larger wavenumbers acquire
additional growth due to the scalaron force with the additional
power $k^{\f{-5+\sqrt{33}}{4(3n+2)}}$ as given in
\eqref{transfer}. 
From \eqref{pwcsmt},
the physical wavenumber of this transition is given by
\be 
\f{k_0}{a_0}=
\left\{
\begin{array}{ll}
1.07\times 10^{-3} h{\rm Mpc}^{-1} &\quad (n=2, \lambda=1) \\
8.44\times 10^{-3} h{\rm Mpc}^{-1} &\quad (n=2, \lambda=3) \\
8.12\times 10^{-2} h{\rm Mpc}^{-1} &\quad (n=2, \lambda=10),
\end{array}
\right.
\ee
that explains the figure well.

In order to make a simple comparison of our results with observations of galaxy 
clustering, we define an effective wavenumber, $\keff(r)$, corresponding to
each length scale $r$, in terms of the top-hat mass fluctuation within the same radius:
\beq
 \sigma^2_r=\int\f{d^3k}{(2\pi)^3}|W(kr)|^2P(k)\equiv
\frac{4\pi \keff^3}{(2\pi)^3}P\mk{\keff(r)},~~~W(kr)\equiv
\f{3j_1(kr)}{kr}.
\eeq
Here $P(k)$ is the linear matter spectrum obtained by the
standard CDM transfer function\cite{Eisenstein:1997ik}
with the scale-invariant initial power spectrum of perturbations, i.e. with the 
primordial spectral index $n_s=1$,
and $W(kr)$ is the Fourier transform of the top-hat window function.

\begin{figure}[t]
\centering
\includegraphics[width=140mm]{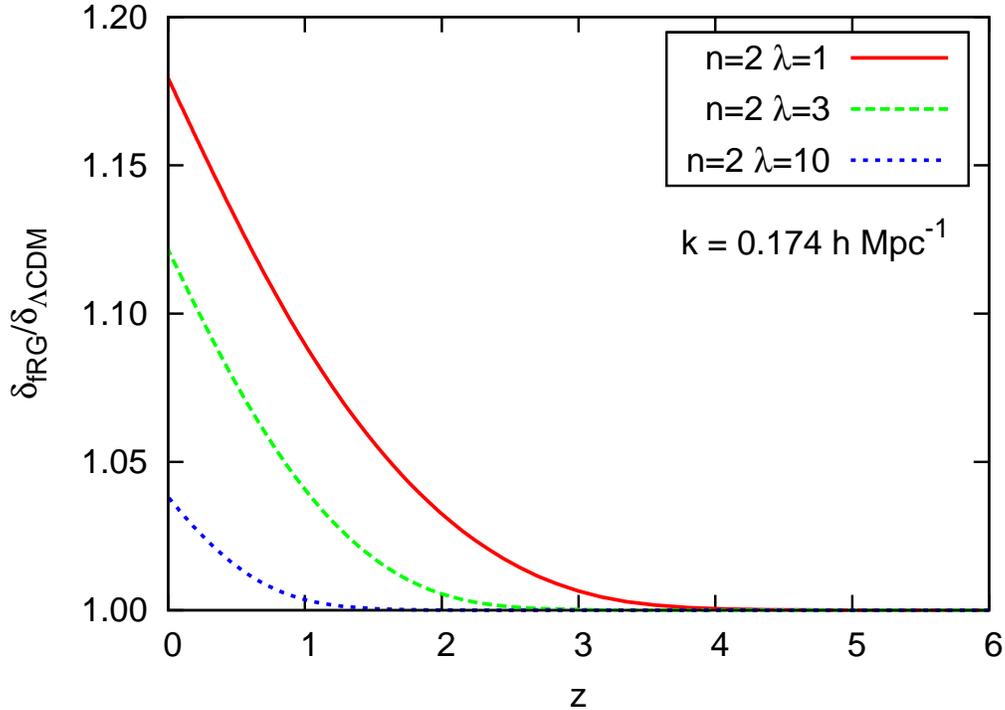}
\caption{The ratio of linear density perturbations $\delta_{\rm fRG}/\delta_{\lcdm}
(k=0.174h{\rm Mpc}^{-1}$) as a function of redshift for three different values
of $\lambda$ with $n=2$.}
\label{fig:drt}
\end{figure}

The wavenumber of our particular interest is the scale
corresponding to $\sigma_8$ normalization, for which we find
$\keff(r=8h^{-1}{\rm Mpc})=0.174h{\rm Mpc}^{-1}$. 
Figure \ref{fig:drt} depicts the redshift evolution of the ratio
$\delta_{\rm fRG}/\delta_{\lcdm}$ for this scale for the same values of
$n$ and $\lambda$ as in Fig.~\ref{fig:dk_n=2_l}. 
Note that this ratio does not stop
growing at the accelerated stage of the Universe expansion which
begins at $z\approx 0.8$ for $\lambda=1$ and $z\approx 0.75$ for
two other values of $\lambda$. 
Since the standard
$\lcdm$ model normalized by large-scale CMB observations explains
galaxy clustering at small scales well, $\delta_{\frg}$ should
not be too much larger than $\delta_\lcdm$ at these scales.  We
may typically require $(\delta_{\frg}/\delta_{\lcdm})^2(k=0.174h{\rm
Mpc}^{-1})\lesssim 1.1$.
Although we neglect non-linear effects here,
the difference between linear calculation and non-linear N-body simulation 
remained smaller than 5\% 
at the wavenumber $0.174h{\rm Mpc}^{-1}$\cite{Oyaizu:2008tb}. 

\begin{figure}[t]
\centering
\subfigure[$n=2$.]{
\includegraphics[width=77mm]{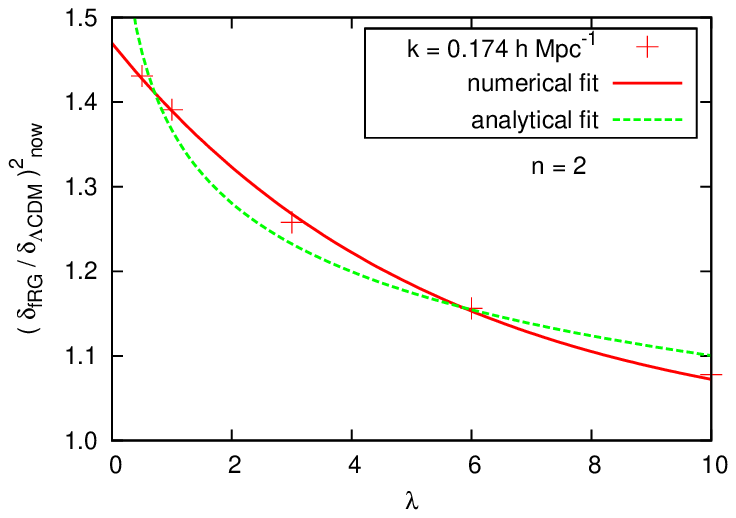}
\label{fig:3a}}
\subfigure[$n=3$.]{
\includegraphics[width=77mm]{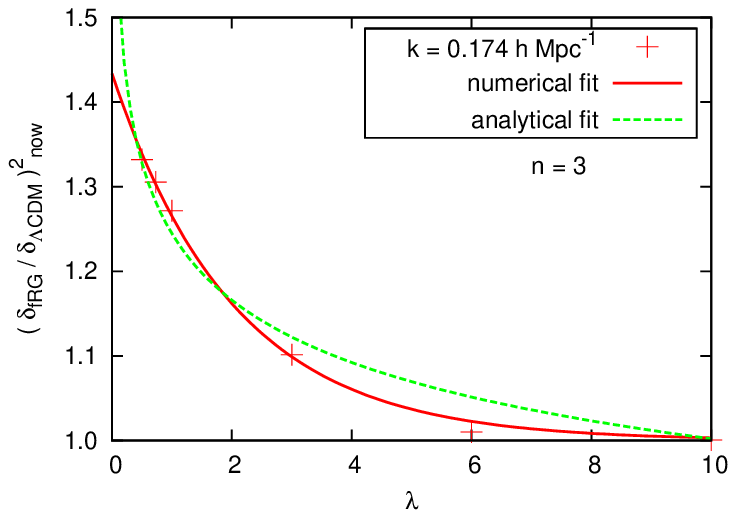}
\label{fig:3b}}
\subfigure[$n=4$.]{
\includegraphics[width=77mm]{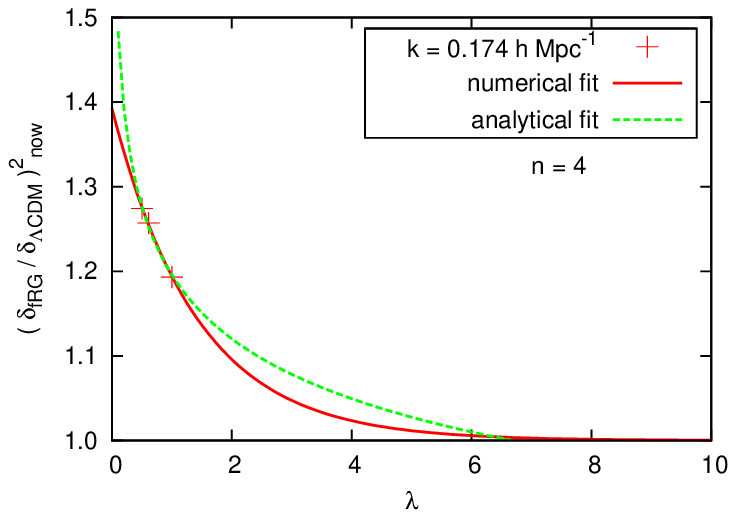}
\label{fig:3c}}
\caption{The present ratio
$(\delta_{\frg}/\delta_{\lcdm})^2(k=0.174h{\rm
Mpc}^{-1})$ as a function of $\lambda$ together with two
fitting functions.}
\label{fig:outpt_w=0.174_rt_p}
\end{figure}

Figures \ref{fig:outpt_w=0.174_rt_p} represent 
$(\delta_{\frg}/\delta_{\lcdm})^2(k=0.174h{\rm Mpc}^{-1})$ 
as a function of $\lambda$ for $n=2,3,$ and 4.
From the analytic formula \eqref{transfer}, 
this $\lambda$ dependence would have the form
$(\delta_{\frg}/\delta_{\lcdm})^2\propto C^2(k)\propto
\lambda^{-\f{(2n-p_n+1)(\sqrt{33}-5)}{4(3n+2)}}$
which is depicted by a broken line in each figure.  
This curve, however,
does not match the asymptotic behaviour $(\delta_{\frg}/\delta_{\lcdm})^2
\longrightarrow 1$ for large $\lambda$.  We find that an
exponential function
\beq
 (\delta_{\frg}/\delta_{\lcdm})^2=1+b_ne^{-q_n\lambda}
\eeq
fits the numerical calculation very well
with $(n, b_n, q_n)=(2, 0.47, 0.19),~ (3, 0.43, 0.49), $ and $(4, 0.39, 0.70)$
, respectively.
From these figures, in order to keep deviation from
$\lcdm$ model smaller than 10\% at $k=0.174h{\rm Mpc}^{-1}$, 
we find $\lambda$ should be larger
than 8.2, 3.0, and 1.9
for $n=2,3,$ and $4$, respectively.

\begin{figure}[t]
\centering
\includegraphics[width=140mm]{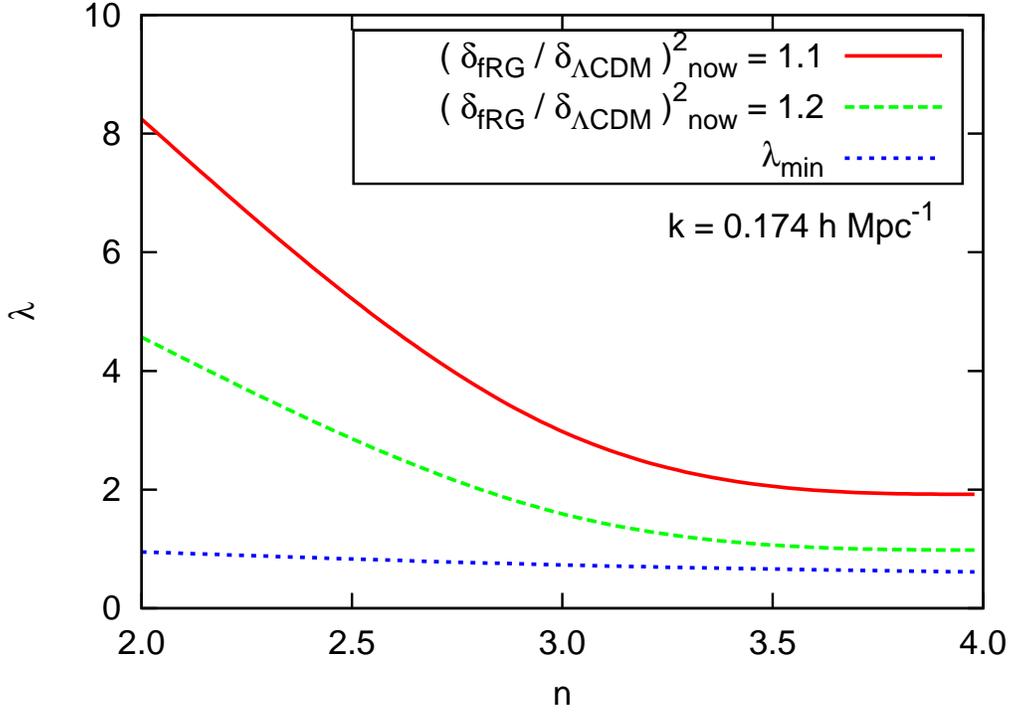}
\caption{Constraints for parameter space.}
\label{fig:n_lam_con}
\end{figure}

From these analysis, we can constrain the parameter space 
as Fig.~\ref{fig:n_lam_con}.
The region which satisfy 
$(\delta_{\frg}/\delta_{\lcdm})^2(k=0.174h{\rm Mpc}^{-1}) < 1.1$ 
corresponds to above the solid line. 
We also show the 20\% boundary by the broken line.
The region below the dotted line is forbidden 
because of instability of the de Sitter regime. 

Next we turn to
another important quantity used to distinguish different theories
of gravity, namely, the gravitational growth index, $\gamma(z)$, of density
fluctuations\cite{Peebles:1984ge,Linder:2005in,Polarski:2007rr,Gannouji:2008wt,
Tsujikawa:2009ku,Narikawa:2009ux}.  It is
defined through
\beq
 \f{d\ln\delta}{d\ln a}=\Omega_m(z)^{\gamma(z)},~~~\text{or}~~~
 \gamma(z)=\f{\log\mk{\f{\dot{\delta}}{H\delta}}}{\log\Omega_m}.
\eeq
It takes a practically constant value $\gamma\cong 0.55$ in the
standard $\lcdm$ model\cite{Peebles:1984ge},
 but it evolves in time in modified
gravity theories in general.  We also note that $\gamma(z)$ has a
nontrivial $k$-dependence in $f(R)$ gravity since density
fluctuations with different wavenumbers evolve differently.
Therefore, this quantity is a useful measure to distinguish
modified gravity from the $\lcdm$ model in the Einstein gravity.

\begin{figure}[t]
\centering
\subfigure[Evolution of $\gamma(z)$ for $n=2$ $\lambda=1$.]{
\includegraphics[width=77mm]{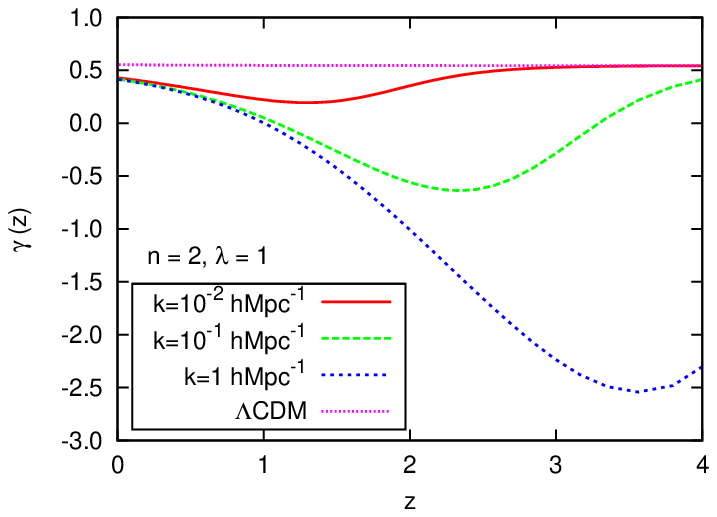}
\label{fig:4a}}
\subfigure[Evolution of $G_{\eff}/G$ for $n=2$ $\lambda=1$.]{
\includegraphics[width=77mm]{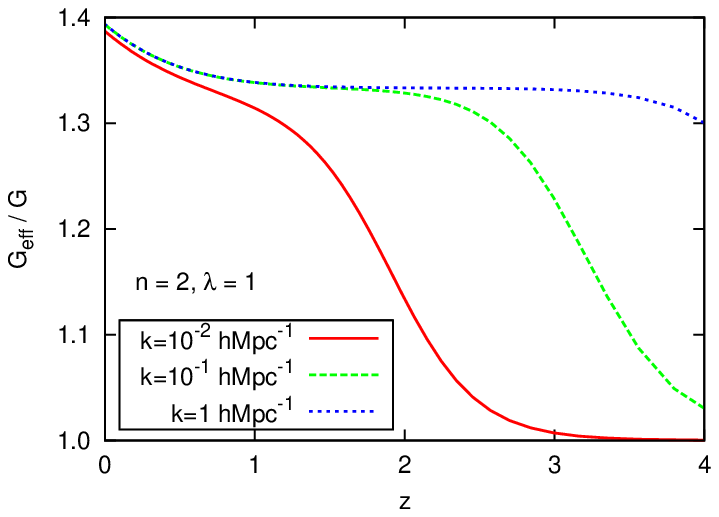}
\label{fig:4b}}
\subfigure[Evolution of $\gamma(z)$ for $n=2$ $\lambda=3$.]{
\includegraphics[width=77mm]{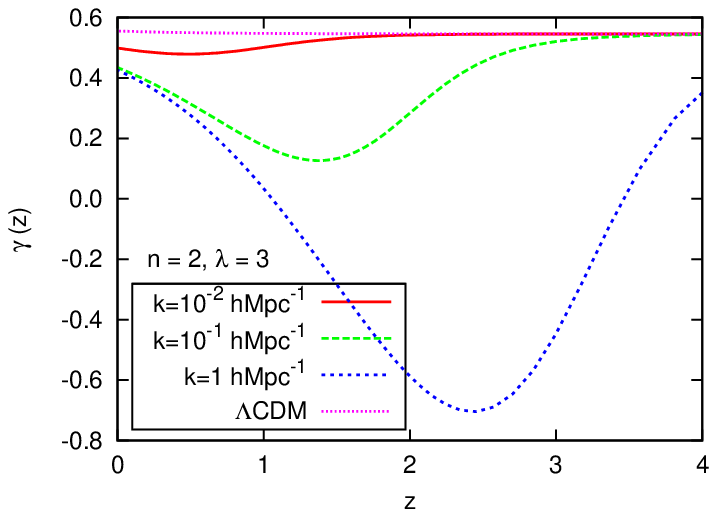}
\label{fig:4c}}
\subfigure[Evolution of $G_{\eff}/G$ for $n=2$ $\lambda=3$.]{
\includegraphics[width=77mm]{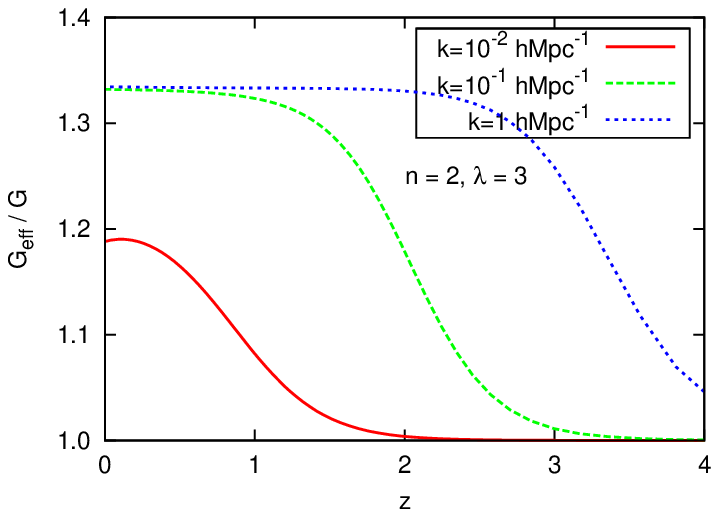}
\label{fig:4d}}
\caption{Evolution of $\gamma(z)$ and $G_{\eff}/G$.}
\label{fig:gG}
\end{figure}

Figures \ref{fig:gG} show evolution of $\gamma(z)$ together with that of
$G_{\eff}/G$ for different values of $k$. 
In the early 
high-redshift regime, $\gamma(z)$ takes a constant value identical
to the $\lcdm$ model because $f(R)$ gravity is indistinguishable
from the Einstein gravity plus a positive cosmological constant then.
It gradually decreases in time, reaches a minimum, and then
increase again towards the present epoch. 
We can understand this
tendency from the evolution equation for $\gamma(z)$\cite{Polarski:2007rr},
\begin{align}
-&(1+z)\ln(1-\Omega_{\rm DE})\f{d\gamma}{dz} \notag \\*
&=-(1-\Omega_{\rm DE})^\gamma
-\f{1}{2}\kk{1+3(2\gamma-1)w_{\rm DE}\Omega_{\rm DE}}
+\f{3}{2}\frac{G_{\eff}}{G}(1-\Omega_{\rm DE})^{1-\gamma},
\end{align}
where $\Omega_{\rm DE}=1-\Omega_m$ is the density parameter of dark energy
based on \eqref{rhoDE}.
In the high-redshift era when $\Omega_{\rm DE}$ is small, the above
equation may be approximated as
\begin{align}
&(1+z)\Omega_{\rm DE}\f{d\gamma}{dz} \notag \\*
&=\f{3}{2}\mk{\f{G_\eff}{G}-1}
+\Omega_{\rm DE}\kk{\f{11}{2}\mk{\gamma-\f{6}{11}}-\f{3}{2}(1-\gamma)\mk{\f{G_\eff}{G}-1}-\f{3}{2}(2\gamma-1)(w_{\rm DE}+1)}.
\end{align}
In the earlier stage, the first term in the right-hand side is more important. 
That explains why $\gamma(z)$ starts to decrease when $G_\eff/G$
starts to increase.  As time goes by towards lower redshifts, the
second term becomes more important to make $\gamma(z)$ increase
again. We note that
recently Narikawa and Yamamoto\cite{Narikawa:2009ux} calculated
time evolution of $\gamma(z)$ in a simplified model \eqref{kinjiF}
numerically and also obtained some analytic expansion, which behaves
qualitatively the same as our numerical results but with much more
exaggerated amplitudes.  Our results, which satisfy all the
viability conditions, exhibit milder deviation from the $\lcdm$ model
than those they found. Existing constraints on the growth index\cite{Rapetti:2009ri} 
are not strong enough to detect any deviation from the $\lcdm$ model and/or to obtain 
new bounds on $f(R)$ DE models, but future observations may reveal its time and wavenumber 
dependence.

\begin{figure}[t]
\centering
\includegraphics[width=140mm]{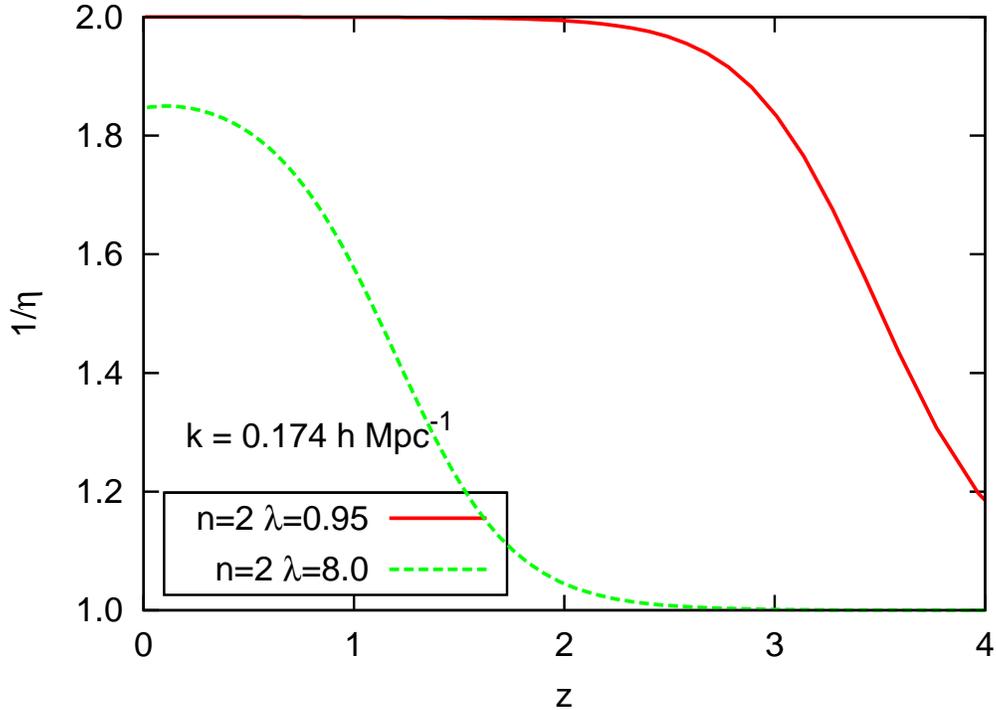}
\caption{The evolution of $1/\eta=\Phi/\Psi$ with $n=2$.}
\label{fig:eta_n=2}
\end{figure}

Another quantity which can characterize the evolution of density perturbations
more directly is the ratio $\delta_{\frg}(z=0.5)/\delta_{\frg}(z=0)$. However, 
it varies only from 0.75 to 0.78 for different choices of the
model parameters when the current matter density parameter is
fixed to $\Omega_{m0}=0.27$ and $n\ge 2$.  This variation is smaller than
that caused by the uncertainty of $\Omega_{m0}$\cite{Gannouji:2008wt}.
So, at present it does not help much to single out the best DE model among the considered ones, 
in contrast to the $f(R)$ DE model\cite{Hu:2007nk} (it has the same behaviour
\eqref{kinjiF} for $R\gg R_s$) in the case corresponding to $n=0.5$ in our notations
which was recently studied in Ref.~\cite{SVH09}.

Finally we consider the quantity $1/\eta=\Phi/\Psi$, 
namely the ratio of gravitational potential to curvature perturbation, 
for which some results from observational data 
were recently obtained in Ref.~\cite{Bean:2009wj}.
In $f(R)$ gravity, $1/\eta$ is expressed as
\be \f{1}{\eta} = 2-\f{1}{1+2\f{k^2}{a^2}\f{F'}{F}}. \ee
Due to the stability conditions $F>0,~F'>0$, this quantity always lies between 1 and 2.
Thus, stable $f(R)$ DE models may not explain such a large value of $1/\eta$ which
is presented in Ref.~\cite{Bean:2009wj} for the redshift interval $1<z<2$.
Figure \ref{fig:eta_n=2} shows the evolution of $1/\eta$ for $n=2$ and 
$\lambda=0.95$ (the minimal possible value) and $8$. 

\section{Conclusions}

In the present paper we have numerically calculated the evolution of
both homogeneous background and density fluctuations in a viable $f(R)$ DE model  
based on the specific functional form proposed in Ref.\ \cite{Starobinsky:2007hu}. 
We have found that viable $f(R)$ gravity models of present DE and 
accelerated expansion of the Universe generically exhibit phantom behaviour during the
matter-dominated stage with crossing of the phantom boundary
$w_{\rm DE}=-1$ at redshifts $z\lesssim 1$. 
The predicted time evolution of 
$w_{\rm DE}$ has qualitatively the same behaviour as that was recently 
obtained from observational data in Ref.~\cite{Shafieloo:2009ti} . 
However, it is important 
that the condition of stability, or even metastability, 
of the future de Sitter epoch strongly restricts possible deviation of $w_{\rm DE}$
from $-1$ by several percents in these models. Thus, the DE phantomness should be
small, if exists at all, that agrees with present observational data. Still
for the models considered, it is not so hopelessly small as in the case of the similar 
model\cite{Hu:2007nk} with $n=0.5$ recently considered in Ref.~\cite{SVH09} using data on 
cluster abundance.   
Note also, that in contrast to Ref.~\cite{BBDS08}, we do not impose
the so called thin-shell condition $|\Delta(f'(R)-1)|\lesssim
|\Phi_N|$, where $\Phi_N$ is the Newtonian potential of matter
inhomogeneities and $\Delta$ means change in the quantity in
question, for scales exceeding galatic ones where a
background matter density approaches the cosmological one.
On the other hand, this condition is satisfied automatically
for matter overdensities more than $10$ for the parameter range
$n\ge 2$ considered in our paper.

As for the density fluctuations, we have numerically confirmed
our previous analytic results of a shift in the power spectrum index
for larger wavenumbers which exceed the scalaron mass during the matter-dominated
epoch\cite{Motohashi:2009qn}, while for smaller wavenumbers
fluctuations have the same amplitude as in the $\lcdm$ model. Once more, the
future de Sitter epoch stability condition bounds possible increase in density
fluctuations for cluster scales (compared to the $\lcdm$ model) by $\sim 20\%$ for $n\ge 2$. 
On the contrary, if it is proven from observational data that
this increase is less than $5\%$, then the background evolution
should be practically indistinguishable from the $\lcdm$
one: $|w_{\rm DE}+1|<10^{-4}$ for $n=2$. This shows that $\sigma_8$
and related density perturbations tests are the most critical
ones for the $f(R)$ DE models considered in the paper.
We have obtained that the upper limit on $|w_{\rm DE}+1|$ for 
$n=2$ and $\lambda = 8$ is $4.4\times 10^{-5}$ when $z=0.16$, 
which is of the same order as $B(0)$.

We have also investigated the growth index $\gamma(k,z)$ of density fluctuations
and have presented an explanation of its anomalous evolution in terms
of time dependence of $G_{\eff}$.  Since $\gamma$ has characteristic
time and wavenumber dependence, future detailed observations may
yield useful information on the validity of $f(R)$ gravity through
this quantity, although current constraints have been obtained
assuming that it is constant both in time and in wavenumber\cite{Bean:2009wj,Rapetti:2009ri}.
Another related observational test of this model is supplied
by the large-scale structure of the Universe which should be
different from that in the $\lcdm$ model. 
In particular, voids are expected to be more pronounced since
the effective gravitational constant is bigger inside them
compared to large matter overdensities where it is practically
equal to that measured in laboratory.

\acknowledgments{
HM and JY are grateful to T.\ Narikawa and K.\ Yamamoto for useful
communications.
We thank T.\ Kobayashi for poiting out
a typo in Ref.~\cite{Motohashi:2009qn} and in the first version of this manuscript.
AS acknowledges RESCEU hospitality as a visiting professor. He was also
partially supported by the grant RFBR 08-02-00923 and by the Scientific
Programme ``Astronomy'' of the Russian Academy of Sciences.
This work was supported in part by
JSPS Grant-in-Aid for Scientific Research No.\ 19340054(JY),
JSPS Core-to-Core program  ``International Research Network
on Dark Energy'', and
Global COE Program ``the Physical Sciences Frontier'', MEXT, Japan.
}

\end{document}